\documentstyle[12pt]{article}
\begin{document}
\thispagestyle{empty}
\begin{center}

\null
\vskip-1truecm
\rightline{IC/95/160}
\rightline{INTERNAL REPORT}
\rightline{(Limited Distribution)}
\vskip1truecm
International Atomic Energy Agency\\
and\\
United Nations Educational Scientific and Cultural Organization\\
\medskip
INTERNATIONAL  FOR THEORETICAL PHYSICS\\
\vskip2truecm
{\bf AN INVESTIGATION OF SINGULAR LAGRANGIANS\\
AS FIELD SYSTEMS\\}
\vskip2truecm
Eqab M. Rabei\footnote{\normalsize Permanent address:
Department of Physics, Mutah University, P.O. Box 7, Karak, Jordan. FAX:
(962--6) 654061}\\
International Centre for Theoretical Physics, Trieste, Italy.\\
\end{center}
\vskip1truecm
\centerline{ABSTRACT}
\baselineskip=20pt
\bigskip

The link between the treatment of singular Lagrangians as field systems
and the
general approach is studied. It is shown that singular Lagrangians as field
systems are always in exact agreement with the general approach.

Two examples and the singular Lagrangian with zero rank Hessian matrix are
studied. The equations of motion in the field systems are equivalent to the
equations which contain acceleration, and the constraints are equivalent to the
equations which do not contain acceleration in the general approach treatment.
\vskip2truecm
\begin{center}
{MIRAMARE -- TRIESTE\\
July 1995\\}
\end{center}

\newpage

\section{INTRODUCTION}
\bigskip

In previous papers [1--4] the Hamilton--Jacobi formulation of singular systems
has been studied. This formulation leads us to the following total differential
equations:
\setcounter{equation}{0}
\renewcommand{\theequation}{1.\arabic{equation}}
\begin{equation}
dq_a={\partial H'_\alpha\over\partial p_a}\ dx_\alpha ,\quad dp_a=-{\partial
H'_\alpha\over\partial q_a}\ dx_\alpha,\quad dp_\alpha =-{\partial
H'_\beta\over\partial x_\alpha}\ dx_\beta
\end{equation}
$$
\alpha ,\beta =0,1,\dots , r;\qquad a=1,2,\dots , n-r\
$$
with constraints
\begin{equation}
H'_\alpha =H_\alpha (x_\beta , q_a,p_a)+p_\alpha
\end{equation}
(Note that we are adopting summation convention in this work.) Solutions of
Eqs.(1.1) give the field, $q_a$, in terms of independent coordinates
\begin{equation}
q_a\equiv q_a(t,x_\mu ),\qquad \mu =1,2,3,\dots ,r
\end{equation}
where $x_0=t$. The link between the Hamilton--Jacobi approach and the Dirac
approach is studied in Ref.[5].

In Ref.[6] the singular Lagrangians are treated as continuous systems. The
Euler--Lagrange equations of constrained systems are proposed in the form
\begin{equation}
{\partial\over\partial x_\alpha}\ \left[ {\partial L'\over \partial
(\partial_\alpha q_a)}\right] -{\partial L'\over\partial q_a}=0;\quad
\partial_\alpha q_a={\partial q_a\over\partial x_\alpha}\ .
\end{equation}
with constraints
\begin{eqnarray}
dG_0&=&-{\partial L'\over\partial t}\ dt\\
dG_\mu &=& -{\partial L'\over\partial x_\mu}\ dt
\end{eqnarray}
where
\begin{equation}
L'(x_\alpha ,\partial_\alpha q_a,\dot x_\mu ,q_a)\equiv L(q_a,x_\alpha ,\dot
q_a=(\partial_\alpha q_a)\dot x_\alpha )
\end{equation}
$$
\dot x_\mu ={dx_\mu\over dt};\qquad \dot x_0=1
$$
\begin{equation}
G_\alpha =H_\alpha \left( q_a,x_\beta ,p_a={\partial L\over\partial \dot
q_a}\right)
\end{equation}
The variation of constraints (1.5,6) should be taken into consideration in
order to have a consistent theory.

An instructive work is the canonical formalism for degenerate Lagrangians
[7--9]; the starting point of this formalism is to consider Lagrangians with
ranks of the Hessian matrix less than $n$. Shanmugadhasan has called these
systems as degenerate. For such systems some of the Euler--Lagrange equations
do not contain acceleration. Following Refs.[7--9] these equations are
considered as constraints. In other words, if the rank of Hessian matrix is
$(n-r)$, with $r<n$, then the Euler--Lagrange equations can be expressed in the
form
\begin{eqnarray}
{d\over dt}\ \left({\partial L\over\partial\dot q_a}\right) -{\partial
L\over\partial q_a} &=& 0\\
{d\over dt}\ \left({\partial L\over\partial\dot x_\mu}\right) -{\partial
L\over\partial x_\mu} &=& 0
\end{eqnarray}

With the aid of Eq.(1.9), Eq.(1.10) can be identically satisfied, i.e. $0=0$,
or they lead to equations free from acceleration. These equations are diveded
into two types: type--$A$ which contains coordinates only and type--$B$ which
contains coordinates and velocities [9]. The total time derivative of the above
two types of constraints should be considered in order to have a consistent
theory.

In this paper we would like to study the link between the treatment of singular
Lagrangians as field systems [6] and the well--known Lagrangian formalism. In
Section 2 the relation between the two approaches is discussed, and in Section
3 two examples of singular Lagrangians are constructed and solved using the two
approaches. In Section 4 the treatment of a singular Lagrangian with Hessian
matrix of zero rank is discussed.

\section{THE RELATION BETWEEN THE TWO APPROACHES}

One should notice that Eqs.(1.4) are equivalent to Eqs.(1.9). In other words
the expressions
\setcounter{equation}{0}
\renewcommand{\theequation}{2.\arabic{equation}}
\begin{equation}
(\partial_\alpha\ q_a)\ \dot x_\alpha
\end{equation}
and
\begin{equation}
\partial_\beta (\partial_\alpha\ q_a\ \dot x_\alpha )\ \dot x_\beta
\end{equation}
In Eqs.(1.4) can be be replaced by $\dot q_a$ and $\ddot q_a$ respectively in
order to obtain Eqs.(1.9). Following Refs.[1--6], we have
\begin{equation}
G_0\equiv H_0\ ,
\end{equation}
and
\begin{equation}
G_\mu\equiv H_\mu =-p_\mu
\end{equation}
Thus, Eqs.(1.6) lead to
\begin{equation}
{dp_\mu\over dt}={\partial L\over\partial x_\mu}
\end{equation}
Making use of the definition of momenta, Eqs.(2.5) lead to Eqs.(1.10). Hence
Eqs.(1.5,6) are equivalent to Eqs.(1.9,10).

\section{EXAMPLES}

The procedure described in Section 2 will be demonstrated by the following
examples.\\ \\
A.\quad Let us consider a Lagrangian of the form
\setcounter{equation}{0}
\renewcommand{\theequation}{3.\arabic{equation}}
\begin{equation}
L={1\over 2}\ \dot q^2_1+\dot q_1\dot q_2+{1\over 2}\ \dot q^2_2+4q_1\dot
q_2+(2q^2_1+4q_1q_2)
\end{equation}
The Euler--Lagrange equations then read as
\begin{eqnarray}
&&\ddot q_1+\ddot q_2-4\dot q_2-4(q_1+q_2)=0\\
&&\ddot q_1+\ddot q_2+4\dot q_1-4q_1=0
\end{eqnarray}
Substituting Eq.(3.2) in Eq.(3.3), gives a $B$--type constraint
\begin{equation}
B_1=\dot q_2+\dot q_1+q_2=0
\end{equation}
For consistent theory, the time derivative of Eq.(3.4) should be equal to zero.
This leads to the new $B$--type constraint
\begin{equation}
B_2=5\dot q_2+4(q_2+q_1)=0
\end{equation}
Taking the time derivative for the new constraints we get a second order
differential equation for $q_2$
\begin{equation}
5\ddot q_2-4q_2=0
\end{equation}
which has the following solution
$$
q_2=A\ e^{{2\over\sqrt 5}\ t} +B\ e^{-{2\over\sqrt 5}\ t}
$$

Now, let us look at this Lagrangian as a field system. Since the rank of the
Hessian matrix is one, the above Lagrangian can be treated as a field system in
the form
\begin{equation}
q_1=q_1(t,q_2);\qquad x_2=q_2
\end{equation}
Thus, the expression
\begin{equation}
\dot q_1={\partial q_1\over\partial t}+{\partial q_1\over\partial q_2}\ \dot
q_2
\end{equation}
can be replaced in Eq.(3.1) to obtain the following modified Lagrangian $L'$:
\begin{eqnarray}
L' &=& {1\over 2}\ \left[{\partial q_1\over\partial t}+{\partial
q_1\over\partial q_2}\ \dot q_2\right]^2+\left[{\partial q_1\over\partial
t}+{\partial q_1\over\partial q_2}\ \dot q_2\right]\ \dot q_2\nonumber\\
&&+{1\over 2}\ \dot q^2_2+4q_1\dot q_2+(2q^2_1+4q_1q_2)
\end{eqnarray}
Making use of Eqs.(1.4), we have
\begin{equation}
{\partial^2q_1\over\partial t^2}+2{\partial^2q_1\over\partial t\ \partial q_2}\
\dot q_2+{\partial q_1\over\partial q_2}\ \ddot q_2+{\partial^2q_1\over\partial
q^2_2}\ \dot q^2_2+\ddot q_2-4\dot q_2-4(q_1+q_2)=0
\end{equation}
Note that we have made the substitution $\alpha =0,2$ and $a=1$, in order to
get the above equation. Making use of Eq.(3.8) and the fact that
\begin{equation}
\ddot q_1={\partial^2q_1\over\partial t^2}+2{\partial^2q_1\over\partial t\
\partial q_2}\ \dot q_2+{\partial q_1\over\partial q_2}\ \ddot
q_2+{\partial^2q_1\over\partial q^2_2}\ \dot q^2_2
\end{equation}
Eq.(3.10) will be the same as Eq.(3.2).

According to Refs.[1--4] the quantity $H_2$ can be calculated as
\begin{equation}
H_2=-(\dot q_1+\dot q_2+4q_1)
\end{equation}
Hence,
\begin{equation}
G_2=-\left[{\partial q_1\over\partial t}+{\partial q_1\over\partial q_2}\ \dot
q_2\right] -\dot q_2-4q_1
\end{equation}
and taking the total differential of Eq.(3.13) one gets,
\begin{eqnarray}
dG_2 &=&-\Biggl\{\left[{\partial^2q_1\over\partial
t^2}+2{\partial^2q_1\over\partial q_2\ \partial t}\ \dot q_2+{\partial^2
q_1\over\partial q^2_2}\ \dot q_2+{\partial q_1\over\partial q_2}\ \ddot
q_2\right] +\Biggr.\nonumber \\
&&\Biggl. +4\left[{\partial q_1\over\partial t} +{\partial q_1\over\partial
q_2}\
\dot q_2\right] +\ddot q_2\Biggr\}\ dt
\end{eqnarray}
Replacing the expression in the first parenthesis from Eq.(3.10) one gets
\begin{equation}
dG_2=-\left\{ 4\dot q_2+4[q_1+q_2]+4\left[{\partial q_1\over\partial
t}+{\partial q_1\over\partial q_2}\ \dot q_2\right]\right\}\ dt
\end{equation}
Making use of Eq.(1.6), one finds
\begin{equation}
dG_2=-4q_1\ dt
\end{equation}
Equating Eq.(3.16) with Eq.(3.15), we have the following constraint
\begin{equation}
F_1=\dot q_2+q_2+{\partial q_1\over\partial t}+{\partial q_1\over\partial q_2}\
\dot q_2=0
\end{equation}
Using the expression (3.8), one observes that this constraint is equivalent to
the $B$--type constraint (3.4).

For a valid theory, the variation of $F_1$ should be zero; thus one gets
\begin{equation}
dF_1=F_2\ dt=0
\end{equation}
where
\begin{equation}
F_2=5\dot q_2+4q_1+4q_2=0
\end{equation}
This is $B_2$ constraint defined in Eq.(3.5).

Again taking the total differential of the new constraint $F_2$, we have
\begin{equation}
dF_2=[5\ddot q_2-4q_2]\ dt=0
\end{equation}
Thus
\begin{equation}
5\ddot q_2-4q_2=0
\end{equation}
This is a second order differential equation for $q_2$ and is the same as
Eq.(3.6). In addition, the function $G_0$ can be evaluated and
\begin{equation}
G_0={1\over 2}\ \left[{\partial q_1\over\partial t}+{\partial q_1\over\partial
q_2}\ \dot q_2\right]^2+{1\over 2}\ \dot q^2_2+\left[{\partial q_1\over\partial
t}+{\partial q_1\over\partial q_2}\ \dot q_2\right]\ \dot q_2-2q^2_1-4q_1q_2
\end{equation}
where
\begin{equation}
dG_0=4\dot q_2\ F_1\ dt
\end{equation}
and this does not lead to any further constraints.\\ \\
B.\quad Consider the Lagrangian of the form
\begin{equation}
L={1\over 2}\ (\dot q^2_1+\dot q^2_2)+\dot q_1\dot q_2+{1\over 2}\
(q^2_1+q^2_2)
\end{equation}
Then, the Euler--Lagrange equations are given as
\begin{eqnarray}
\ddot q_1+\ddot q_2-q_1 & = & 0\\
\ddot q_1+\ddot q_2-q_2 & = & 0
\end{eqnarray}
Expressing Eq.(3.25) as
\begin{equation}
q_1=\ddot q_1+\ddot q_2
\end{equation}
and substituting in Eq.(3.26), one gets an $A$--type constraint
\begin{equation}
A_1=q_1-q_2=0
\end{equation}

There are no further constraints. Thus Eq.(3.26) takes the form
\begin{equation}
2\ddot q_2-q_2=0
\end{equation}

As in the previous example, this system can be treated as field system, and the
modified Lagrangian $L'$ reads as
\begin{equation}
L'={1\over 2}\left[{\partial q_1\over\partial t}+{\partial q_1\over\partial
q_2}\ \dot q_2\right]^2+{1\over 2}\ \dot q^2_2+\left[{\partial q_1\over\partial
t}+{\partial q_1\over\partial q_2}\ \dot q_2\right]\ \dot q_2+{1\over 2}\
(q^2_1+q^2_2)
\end{equation}
The Euler--Lagrange equation for this field system is obtained as
\begin{equation}
{\partial^2q_1\over\partial t^2}+2{\partial^2q_1\over\partial t\ \partial
q_2}\ \dot q_2+{\partial q_1\over\partial q_2}\ \ddot
q_2+{\partial^2q_1\over\partial q^2_2}\ \dot q^2_2+\ddot q_2-q_1=0
\end{equation}
Again replacing $\ddot q_1$ by the expression (3.11), Eq.(3.31) will be the
same as Eq.(3.25). Besides, the function $G_2$ can be calculated as
\begin{equation}
G_2=-\left[{\partial q_1\over\partial t}+{\partial q_1\over\partial q_2}\ \dot
q_2\right] -\dot q_2
\end{equation}
and the total differential of $G_2$ can be written as
\begin{equation}
dG_2=-\left[{\partial^2q_1\over\partial t^2}+2{\partial^2q_1\over\partial t\
\partial q_2}\ \dot q_2+{\partial^2q_1\over\partial\dot q^2_2}\ \dot
q^2_2+{\partial q_1\over\partial q_2}\ \ddot q_2+\ddot q_2\right]\ dt
\end{equation}
Using Eq.(3.31) and Eq.(1.6), we have
\begin{equation}
dG_2=-q_1\ dt=-q_2\ dt
\end{equation}
which leads to the following constraint
\begin{equation}
F_1=q_1-q_2=0
\end{equation}
This is an $A$--type constraint of the form (3.28).

Taking the total differential of $F_1$, we have
\begin{equation}
dF_1=\left[{\partial q_1\over\partial t}+{\partial q_1\over\partial q_2}\ \dot
q_2-\dot q_2\right]\ dt=0
\end{equation}
and this leads to a new constraint
\begin{equation}
F_2={\partial q_1\over\partial t}+{\partial q_1\over\partial q_2}\ \dot
q_2-\dot q_2=0
\end{equation}
which is equivalent to the total time derivative of the constraint (3.28).

Again calculating the total differential of $F_2$, one gets
\begin{equation}
dF_2=\left[{\partial^2q_1\over\partial t^2}+2{\partial^2q_1\over\partial t\
\partial q_2}\ \dot q_2+{\partial^2q_1\over\partial q^2_2}\ \dot
q^2_2+{\partial q_1\over\partial q_2}\ \ddot q_2-\ddot q_2\right]\ dt=0
\end{equation}
and making use of Eqs.(3.31) and (3.35), we get
\begin{equation}
2\ddot q_2-q_2=0
\end{equation}
which is the same as Eq.(3.29). Besides the function $G_0$ is calculated as
\begin{equation}
G_0={1\over 2}\left[{\partial q_1\over\partial t}+{\partial q_1\over\partial
q_2}\ \dot q_2\right]^2+{1\over 2}\ \dot q^2_2+\left[{\partial q_1\over\partial
t}+{\partial q_1\over\partial q_2}\ \dot q_2\right]\ \dot q_2-{1\over 2}\
[q^2_1+q^2_2]
\end{equation}
Thus, the total differential of $G_0$ is obtained as
\begin{equation}
dG_0=\dot q_2\ F_1\ dt=0
\end{equation}
and with the aid of Eq.(3.35), it is identically satisfied.

\section{A SINGULAR LAGRANGIAN WITH ZERO RANK HESSIAN MATRIX}

According to the treatment of singular Lagrangians as field systems: if the
Hessian matrix has rank equal to zero, the Lagrangian cannot be treated as
field system. Whereas, the equation of motion which does not contain
acceleration can be obtained using the constraints (1.6).

As an example let us consider the following Lagrangian which was given in
Ref.[10]:
\setcounter{equation}{0}
\renewcommand{\theequation}{4.\arabic{equation}}
\begin{equation}
L=(q_2+q_3)\ \dot q_1+q_4\ \dot q_3+{1\over 2}\ (q^2_4-2q_2q_3-q^2_3)
\end{equation}
The momenta are obtained as
\begin{equation}
p_1=q_2+q_3,\quad p_2=0,\quad p_3=q_4,\quad p_4=0
\end{equation}
Thus,
\begin{eqnarray}
G_1 &=& -(q_2+q_3)\\
G_2 &=& 0\\
G_3 &=& -q_4\\
G_4 &=& 0
\end{eqnarray}
Making use of Eqs.(1.6), one gets
\begin{eqnarray}
dG_1 &=& -(\dot q_2+\dot q_3)\ dt=0\\
dG_2 &=& 0=-(\dot q_1-q_3)\ dt\\
dG_3 &=& -\dot q_4\ dt =-(\dot q_1-q_2-q_3)\ dt\\
dG_4 &=& 0=-(\dot q_3+q_4)\ dt
\end{eqnarray}
These equations lead to the following equations of motion
\begin{eqnarray}
\dot q_2+\dot q_3 &=& 0\\
\dot q_1-q_3 &=& 0\\
\dot q_4-\dot q_1+q_2+q_3 &=& 0\\
\dot q_3+q_4 &=& 0
\end{eqnarray}
and these are the Euler--Lagrange equations which are free from acceleration,
and are of $B$--type constraints.

\section{CONCLUSIONS}

As it was mentioned in the introduction if the rank of the Hessian matrix for
discrete systems is $(n-r); 0<r<n$, the systems can be treated as field
systems. It can be observed that the treatment of Lagrangians as field systems
is always in exact agreement with the general approach. The equations of motion
(1.4) are equivalent to the equations of motion (1.9). Besides, the constraints
(1.6) are equivalent to the equations (1.10).

The consistent theory in the treatment of Lagrangians as field systems also
leads to two types of constraints: a $B$--type which contains at least one
member of the set $\left\{\dot q_\mu ,{\partial q_a\over\partial t},{\partial
q_a\over\partial q_\mu}\right\}$, and an $A$--type which contains coordinates
only. As we have seen, in the first example $F_1$ and $F_2$ are $B$--types;
while the constraint $F_1$ in the second example is an $A$--type.

In the general approach the constraints can be obtained from the
Euler--Lagrange
equations, whereas, in the treatment of Lagrangians as field systems, the
constraints can be determined from the relations (1.5,6) and the new
constraints can be obtained using the variations of these relations.
\vspace{1cm}

\noindent{\bf Acknowledgments}

The author would like to thank the International Centre for Theoretical
Physics, Trieste, for hospitality.


\newpage

\begin{thebibliography}{99}

\bibitem{}
Y. G\"uler, Nuovo Cimento B {\bf 107}, 1389 (1992).

\bibitem{}
Y. G\"uler, Nuovo Cimento B {\bf 107}, 1143 (1992).

\bibitem{}
Eqab M. Rabei and Y. G\"uler, Phys. Rev. A {\bf 46}, 3513 (1992).

\bibitem{}
Eqab M. Rabei and Y. G\"uler, Turkish J. Phys. {\bf 16}, 297 (1992).

\bibitem{}
Eqab M. Rabei, On Hamiltonian Systems with constraints, to appear in Hadronic
Journal, Vol.18 (1995).

\bibitem{}
N.I. Farahat and Y. G\"uler, Phys. Rev. A {\bf 51}, 68 (1995).

\bibitem{}
S. Shanmugadhasan, Proc. Camb. Phil. Soc. {\bf 59}, 743 (1963).

\bibitem{}
S. Shanmugadhasan, J. Math. Phys. {\bf 14}, 677 (1973).

\bibitem{}
K. Sundermeyer, Lecture Notes in Physics, Springer--Verlag, (1982).

\bibitem{}
Usha Kulshreshtha, J. Math. Phys. {\bf 33}, 633 (1992).
\end{thebibliography}
\end{document}